\documentclass[aps,pra,longbibliography,twocolumn,showpacs,superscriptaddress,floatfix]{revtex4-2}
\usepackage[utf8]{inputenc} %useful to type directly diacritic characters
\usepackage{amssymb,amsmath}
\usepackage{graphicx}
\usepackage{braket}
\usepackage{siunitx}
\usepackage{xcolor}
\usepackage{romannum}
\usepackage{xparse}
\usepackage[english]{babel}
\usepackage[hidelinks]{hyperref}
\hypersetup{
     colorlinks   = true,
     citecolor    = blue,
     linkcolor     = blue,
     urlcolor      = blue
}
\def\kr{k_R}                            				% kr
\def\kl{k_L}                            				% k\L
\def\Rb87{^{87}\mathrm{Rb}}                     % Rb 87
\def\K40{^{40}\mathrm{K}}                    		% K 40

\def\ex{\mathbf{e}_x}  
\def\es{\mathbf{e}_{\rm s}}

\def\PhiL{\Phi_{\rm L}}
 
\def\HAA{\hat H_{\rm AA}} 
\def\UAA{\hat U_{\rm AA}}

\def\Q{Q_{\rm r}}
\def\P{P_{\rm r}}
\def\mi{m_{\rm i}}
\def\mf{m_{\rm f}}%final

\def\qi{q_{\rm i}}%initial
\def\Phir{\Phi_{\rm r}}%rational

\makeatletter
\newcommand{\phibar}{{\mathchoice
  {\smash@bar\textfont\displaystyle{0.0}{2.5}\phi}
  {\smash@bar\textfont\textstyle{0.0}{2.5}\phi}
  {\smash@bar\scriptfont\scriptstyle{0.0}{2.5}\phi}
  {\smash@bar\scriptscriptfont\scriptscriptstyle{0.0}{2.5}\phi}
}}
\newcommand{\Phibar}{{\mathchoice
  {\smash@bar\textfont\displaystyle{0.55}{2.0}\Phi}
  {\smash@bar\textfont\textstyle{0.55}{2.0}\Phi}
  {\smash@bar\scriptfont\scriptstyle{0.55}{2.0}\Phi}
  {\smash@bar\scriptscriptfont\scriptscriptstyle{0.55}{2.0}\Phi}
}}
\newcommand{\qbar}{{\mathchoice
  {\smash@bar\textfont\displaystyle{1.5}{0.5}q}
  {\smash@bar\textfont\textstyle{1.5}{0.5}q}
  {\smash@bar\scriptfont\scriptstyle{1.5}{0.5}q}
  {\smash@bar\scriptscriptfont\scriptscriptstyle{1.5}{0.5}q}
}}
\newcommand{\smash@bar}[4]{%
  \smash{\rlap{\raisebox{-#3\fontdimen5#10}{$\m@th#2\mkern#4mu\mathchar'26$}}}%
}
\newcommand{\RNum}[1]{\uppercase\expandafter{\romannumeral #1\relax}}

\ExplSyntaxOn
\NewDocumentCommand{\mref}{m}{\quinn_mref:n {#1}}
\seq_new:N \l_quinn_mref_seq
\cs_new:Npn \quinn_mref:n #1
 {
  \seq_set_split:Nnn \l_quinn_mref_seq { , } { #1 }
  \seq_pop_right:NN \l_quinn_mref_seq \l_tmpa_tl
  ( % print the left parenthesis
  \seq_map_inline:Nn \l_quinn_mref_seq
    { \ref{##1},\nobreakspace } % print the first references
  \exp_args:NV \ref \l_tmpa_tl % print the last or only one
  ) % print the right parenthesis
 }
\ExplSyntaxOff

\makeatother        

\begin{document}
\pagenumbering{arabic}
%Title of paper
\title{Coherence and decoherence in the Harper-Hofstadter model}
%Coherence in a closed quantum system with the Harper-Hofstadter model in a tube-shape geometry

%Coherence and decoherence with the quasi-crystalline Harper-Hofstadter model

% language tip: in xx geometry
% e.g. Band gap closing in a synthetic Hall tube of neutral fermions

\author{Q.-Y.~Liang}
\affiliation{Joint Quantum Institute, National Institute of Standards and Technology, and University of Maryland, Gaithersburg, Maryland, 20899, USA}
%Dimitrios Trypogeorgos
\author{D.~Trypogeorgos}
\affiliation{CNR Nanotec, Institute of Nanotechnology, via Monteroni, 73100, Lecce, Italy}
%Ana Vald\'{e}s-Curiel
\author{A.~Vald\'{e}s-Curiel}
\affiliation{Joint Quantum Institute, National Institute of Standards and Technology, and University of Maryland, Gaithersburg, Maryland, 20899, USA}

\author{J. Tao}
\affiliation{Joint Quantum Institute, National Institute of Standards and Technology, and University of Maryland, Gaithersburg, Maryland, 20899, USA}
\author{M. Zhao}
\affiliation{Joint Quantum Institute, National Institute of Standards and Technology, and University of Maryland, Gaithersburg, Maryland, 20899, USA}
\author{I.~B.~Spielman}
\affiliation{Joint Quantum Institute, National Institute of Standards and Technology, and University of Maryland, Gaithersburg, Maryland, 20899, USA}
\email{ian.spielman@nist.gov}
\homepage{http://ultracold.jqi.umd.edu}
\date{\today}
\begin{abstract}
We quantum-simulated the 2D Harper-Hofstadter (HH) lattice model in a highly elongated tube geometry---three sites in circumference---using an atomic Bose-Einstein condensate. 
In addition to the usual transverse (out-of-plane) magnetic flux, piercing the surface of the tube, we threaded a longitudinal flux $\PhiL$ down the axis of the tube.
This geometry evokes an Aharonov-Bohm interferometer, where noise in $\PhiL$ would readily decohere the interference present in trajectories encircling the tube.
We observe this behavior only when transverse flux is a rational fraction of the flux-quantum, and remarkably find that for irrational fractions the decoherence is absent.
Furthermore, at  rational values of transverse flux, we show that the time evolution averaged over the noisy longitudinal flux matches the time evolution at nearby irrational fluxes.
Thus, the appealing intuitive picture of an Aharonov-Bohm interferometer is insufficient. Instead, we quantitatively explain our observations by transforming the HH model into a collection of momentum-space Aubry-Andr\'{e} models.
\end{abstract}

\maketitle

% \clearpage

Understanding how and when closed quantum systems lose or retain coherence is a central intellectual and practical question for quantum technologies. 
For example, modern optical atomic clocks operate in highly optimized decoherence-free subspaces created by using ``clock’’ states~\cite{Ye2008} that are insensitive to the environment, as well as using lasers at magic wavelengths and polarizations that give only common mode energy shifts. 
In rare cases, such as collisional narrowing~\cite{Dicke1953} or environment assisted tunneling~\cite{Maier2019}, random processes can enhance coherence.
Here we add to this list the quasi-periodic lattice described by the Harper-Hofstadter (HH) model~\cite{Harper1955,Hofstadter1976} in a highly-elongated tube geometry---a 1D quasi-crystal---by showing that the dynamics can be made immune to environmental noise.

Ultracold atomic gases in optical lattices can mimic Aharonov-Bohm (AB) phase factors using the optical phase of interfering laser beams~\cite{goldman2016topological,Cooper2019}.
Even in units of the magnetic flux quantum $\Phi_0= h/q$, these systems realize large tunable magnetic fluxes $\Phi=a^2 B/\Phi_0$ per lattice plaquette, where $h$, $q$, $a$ and $B$ denote Planck's constant, charge, lattice constant and a uniform magnetic field, respectively.
Planar geometries~\cite{miyake2013realizing,aidelsburger2013realization,dean2013hofstadter}, narrow Hall ribbons~\cite{stuhl2015visualizing,mancini2015observation} and even tubes~\cite{han2019band,Li2018} have been realized in experiment.
In the tube geometry, the longitudinal flux $\PhiL$ threading the tube has significant physical consequences~\cite{hainaut2018controlling,luo2020tunable}. 
For example, adiabatically ramping $\PhiL$ by one flux quantum would drive one cycle of Laughlin's topological charge pump~\cite{laughlin1981quantized} that can probe both non-interacting and many-body topological systems~\cite{taddia2017topological}.

\begin{figure}[ht!]
	\begin{center}
		\includegraphics{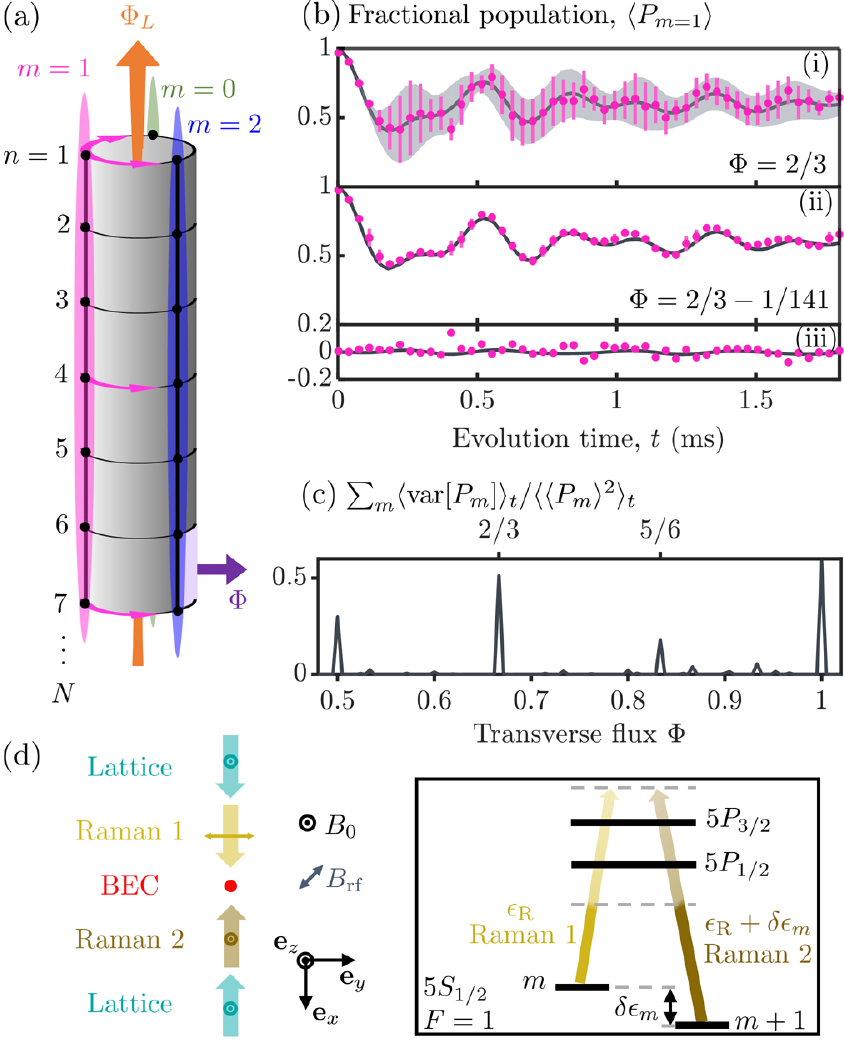}
		\end{center}
	\caption{\textbf{Schematic and summary of results} 
({a}) The HH model in the tube geometry.
({b}) Average fractional population $\langle P_{m=1}\rangle$ at $\Phi=2/3$ (\romannum{1}) and $\Phi=2/3-1/141$ (\romannum{2}), along with their difference (\romannum{3}).
The pink uncertainty bars denote the standard deviation from the mean across 14 to 22 experimental runs and result from the sensitivity to $\phi$ along with technical noise from: variations in magnetic field, Raman coupling strength, and rf field strength, along with statistical detection uncertainties. 
The dark blue curves and error bands plot our numerical simulations.
({c}) Numerically simulated phase noise sensitivity quantified by the normalized variance time-averaged over the range $[0,1.8]\ {\rm ms}$. 
({d}) Raman coupling scheme and schematic of experimental setup.
The optical lattice beam was retro-reflected forming a standing wave.
The Raman beams were orthogonally polarized and colinear with the lattice beams. 
%Raman 2 contains three frequency components $\epsilon_{\rm R}+\delta\epsilon_{m}$, $m=0,1,2$, to realize the cylindrical geometry, while $\epsilon_{\rm{R}}+\delta\epsilon_2$ is removed to realize the planar geometry.
}  
	\label{fig1}
\end{figure}

The HH model was initially formulated to describe electrons moving in a 2D crystalline lattice with a transverse magnetic field; in terms of the AB phase $\Phi$ the HH model is
\begin{align}
\hat{H}=&-J_{\rm s}\sum_{m,n}e^{i(2\pi\Phi n+\phi)}\ket{m+1,n}\bra{m,n}\nonumber\\
&-J_x\sum_{m,n}\ket{m,n+1}\bra{m,n}+{\rm h.c.}.
\label{eq1}
\end{align}
For isotropic tunneling $J_{\rm s}=J_x$, the resulting ``Hofstadter butterfly'' energy spectrum was one of the first quantum fractals ever predicted~\cite{Hofstadter1976}.
In planar geometries a uniform Peierls phase $\phi$ has no physical consequence; however, for a tube $M$ sites in circumference [$M=3$ depicted in Fig.~\ref{fig1}(a)], the uniform Peierls phase contributes $M \phi/(2\pi)$ to the longitudinal flux $\PhiL$~\footnote{Summing the phase factors around the tube gives $\PhiL=M[\Phi n+\phi/(2\pi)]$; $\PhiL$ depends on longitudinal position $n$ owing to the contribution $2\pi i\Phi n$ from the transverse flux.}.
Using the synthetic dimension approach~\cite{stuhl2015visualizing,mancini2015observation,Meier2018}, we assembled our 2D lattice by combining the sites of a 1D optical lattice with three internal atomic states to  respectively define the longitudinal ($\ex$) and azimuthal ($\es$) directions of our tube.
We performed interference experiments [Fig.~\ref{fig1}(a)] akin to AB interferometers: particles prepared at site $m=1$ along $\es$, but extended along $\ex$, were released and potentially interfered as they rapidly encircled the tube.

In this Letter, we report three key observations summarized in Fig.~\ref{fig1}(b), where $\langle\cdot\rangle$ denotes the average over $\phi\in[0,2\pi)$, and $\langle\cdot\rangle_t$ marks the time-average.
(i) For rational transverse flux $\Phi=P/Q$ (expressed in reduced form), the time-evolving population in each $m$-site depends strongly on $\phi$ and therefore exhibits large uncertainties, as one would expect for an AB interferometer.
(ii) This dependence decreases with increasing $Q$ and vanishes for irrational $\Phi$.
(iii) The $\phi$-averaged dynamics at rational $\Phi$ are equal to those at nearby irrational $\Phi$.
In all cases, our numerical simulations are in excellent agreement our observations.
While we experimentally probed $\Phi$ near 2/3, these observations are generalized by the numerical simulation shown in Fig.~\ref{fig1}(c) that plots the time-averaged sensitivity to $\phi$; this curve is approximated by an everywhere discontinuous Thomae-like function~\footnote{The Thomae function is zero for irrational $\Phi$ and $1/Q$ for rational $\Phi$.  Our analysis [Eq.~\eqref{eq: Thomae function}] shows that the expected structure is proportional to a modified Thomae function that becomes $1/{\rm LCM(Q,M)}$ for rational $\Phi$.}.
Our system's spatial extent $w$ limits the degree to which $\Phi$ can be distinguished to $\approx a/w$, broadening the otherwise singular peaks.

%%%%%%%%%%%%%%%%%%%%%%%%%%%
% Experimental setup, cycle and detection 
%%%%%%%%%%%%%%%%%%%%%%%%%
% Beam geometry
% wavelengths
% Relate the fluxes to the wavevectors
% Laser sequence
%%%%%%%%%%%%%%%%%%%%%%%%%%%%

% TF radius from TOF 11.6±0.5
% TF radius from in-situ 11.4±0.6
% along the lattice direction
% 11.5/0.532*2=43.2

{\it Implementation} We performed these experiments using $\textsuperscript{87}$Rb BECs in the $\ket{5S_{1/2},F=1}$ electronic ground state manifold in a crossed optical dipole trap~\cite{Lin2009}.
The longitudinal Thomas-Fermi radius $R_{\rm TF}=11.5(5)\ \mu {\rm m}$ was obtained both by direct in-situ imaging and mean-field driven expansion~\cite{castin1996bose}.
We used continuous dynamical decoupling (CDD) to eliminate the magnetic field sensitivity~\cite{trypogeorgos2018synthetic, anderson2020realization} of this manifold's three $\ket{m_F}$ states, giving three dressed $\ket{m}$ states that served as sites along $\es$, with $m=0,1,2$.
Our experiments took place in a $B_0\approx31.47\ {\rm G}$ bias field, along with a resonant $22.1\ {\rm MHz}$ radio-frequency (rf) magnetic field $B_{\rm rf}$ with $150\ {\rm kHz}$ Rabi frequency for CDD.
The resulting energy differences $\delta\epsilon_{m}\equiv \epsilon_m-\epsilon_{m+1}$ in the CDD basis were $(\delta\epsilon_{0}, \delta\epsilon_{1}, \delta\epsilon_{2})=h\times(-308.3,118.6,189.7)\ {\rm kHz}$.

\begin{figure}[t!]
	\begin{center}
		\includegraphics{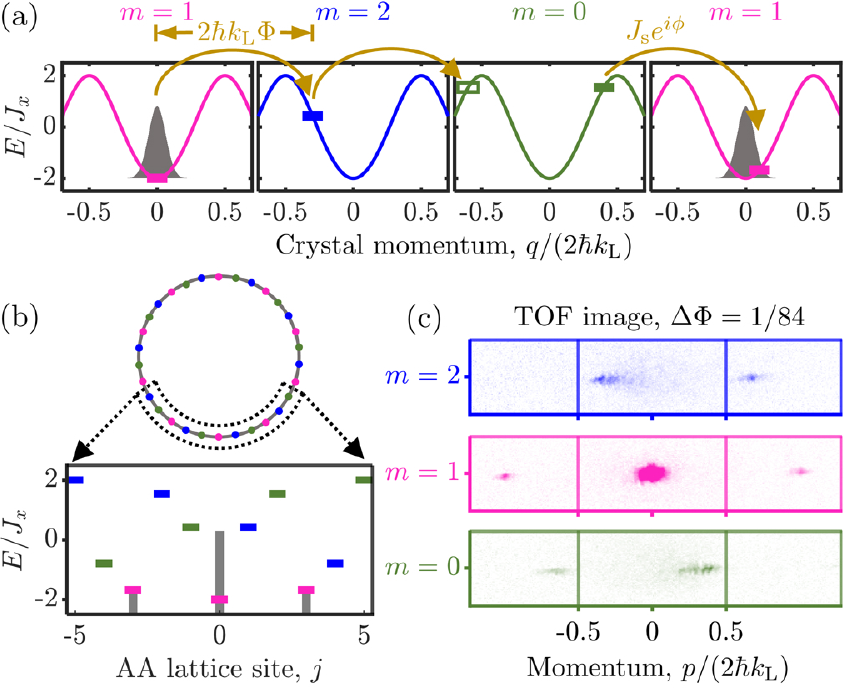}
		\end{center}
	\caption{\textbf{Momentum space AA model.}
({a}) Lowest band of the longitudinal lattice.
The filled grey curve represents a Gaussian wavepacket of width $0.064\times 2\hbar k_{\rm L}$. The hollow and solid green rectangles, horizontally spaced by $2 \hbar k_{\rm L}$, are equivalent points.
The arrows mark Raman-induced coupling starting in $\ket{j=0}=\ket{m_0=1,q_0=0}$ with $\Delta\Phi=1/30$.
({b}) AA ring corresponding to (a). 
The bottom panel zooms into the section $-5\leq j\leq5$.
The rectangles marking the on-site energy resulting from sampling the dispersion in (a), are colored in accordance to their $\ket{m}$ state, and the grey bars indicate the discretely sampled Gaussian wavepacket.
({c}) TOF data taken at $t=1.506\ {\rm ms}$ show the momentum distribution associated with each $\ket{m}$ site.
Each crystal momentum state of the longitudinal lattice consists of momentum states that are imaged as horizontally spaced diffraction orders with spacing $2 \hbar k_{\rm L}$.
The synthetic lattice sites are resolved vertically; the diffraction orders in these sites are shifted by $2 \hbar k_{\rm L}\Phi$ for each synthetic hopping event as expected from (a).
}

	\label{fig2} 
\end{figure} 

A retro-reflected $\lambda_{\rm L}=532.008(5)\ {\rm nm}$ laser beam along ${\bf e}_x$ defined our tube's longitudinal sites $\ket{n}$ with spacing $a=\lambda_{\rm L}/2$, and the $V=5.0(1)E_{\rm L}$ lattice depth set the tunneling matrix element $J_x = 0.066(2) E_{\rm L}$.
Here, $E_{\rm L}=\hbar^2 k_{\rm L}^2/(2 M_a )$ and $k_{\rm L} = 2\pi / \lambda_{\rm L}$ are the single-photon recoil energy and wavevector, respectively, for atoms with mass $M_a$.
%A pair of Raman laser beams [Fig.~\ref{fig1}(d)], with wavelength $\lambda_{\rm R}$, and recoil wavevector $k_{\rm R} = 2\pi / \lambda_{\rm R}$, resonantly coupled the $\ket{m}$ states with strength $\Omega_{\rm R} = 0.296(6)E_{\rm L}$, providing hopping $J_{\rm s}=0.111(3)E_{\rm L}$ along $\es$.
A pair of Raman laser beams [Fig.~\ref{fig1}(d)] resonantly coupled the $\ket{m}$ states with equal strength $\Omega_{\rm R} = 0.296(6)E_{\rm L}$, providing hopping $J_{\rm s}=0.111(3)E_{\rm L}$ along $\es$.
Raman 2 contained three frequency components $\epsilon_{\rm R}+\delta\epsilon_{m}$ in the CDD basis to realize the cylindrical geometry, and $\epsilon_{\rm{R}}+\delta\epsilon_2$ was removed to realize a planar geometry.
Because $\ket{m=2}$ was Raman-coupled to $\ket{m=0}$, we adopt periodic labels, i.e., $\ket{m}\equiv\ket{{\rm mod}(m,3)}$.
Since $\Phi=k_{\rm R}/k_{\rm L}$ \footnote{The position-space Raman matrix elements $\propto e^{i(2k_{\rm R} x +\phi)} \ket{m+1, x}\bra{m,x}$, sampled on a lattice with positions $x = n a$, yield the phase factor in Eq.~\eqref{eq1}, with $\Phi=k_{\rm R}/k_{\rm L}$.}, with Raman recoil wavevector $k_{\rm R} = 2\pi / \lambda_{\rm R}$, we tuned $\Phi = 2/3 + \Delta\Phi$ by varying the Raman wavelength $\lambda_{\rm R}$ from $770.94(1)\ {\rm nm}$ ($\Delta\Phi=2/87$) to $806.46(1)\ {\rm  nm}$ ($\Delta\Phi=-1/141$). The tuning range was limited by the increasing power requirement as the detuning from the excited states increased.
%While $\phi$ could be tuned by changing the phase of any of the Raman lasers or displacing the optical lattice, we instead phase-shifted the CDD field $B_{\rm rf}$, giving the same effect in the CDD basis~\cite{Campbell2016a}.
Each experimental run randomly sampled a Peierls phase $\phi$ uniformly distributed from 0 to $2 \pi$, by phase-shifting the CDD rf field $B_{\rm rf}$~\cite{Campbell2016a}.

%\onecolumngrid

\begin{figure*}[t!]
	\begin{centering}
		\includegraphics{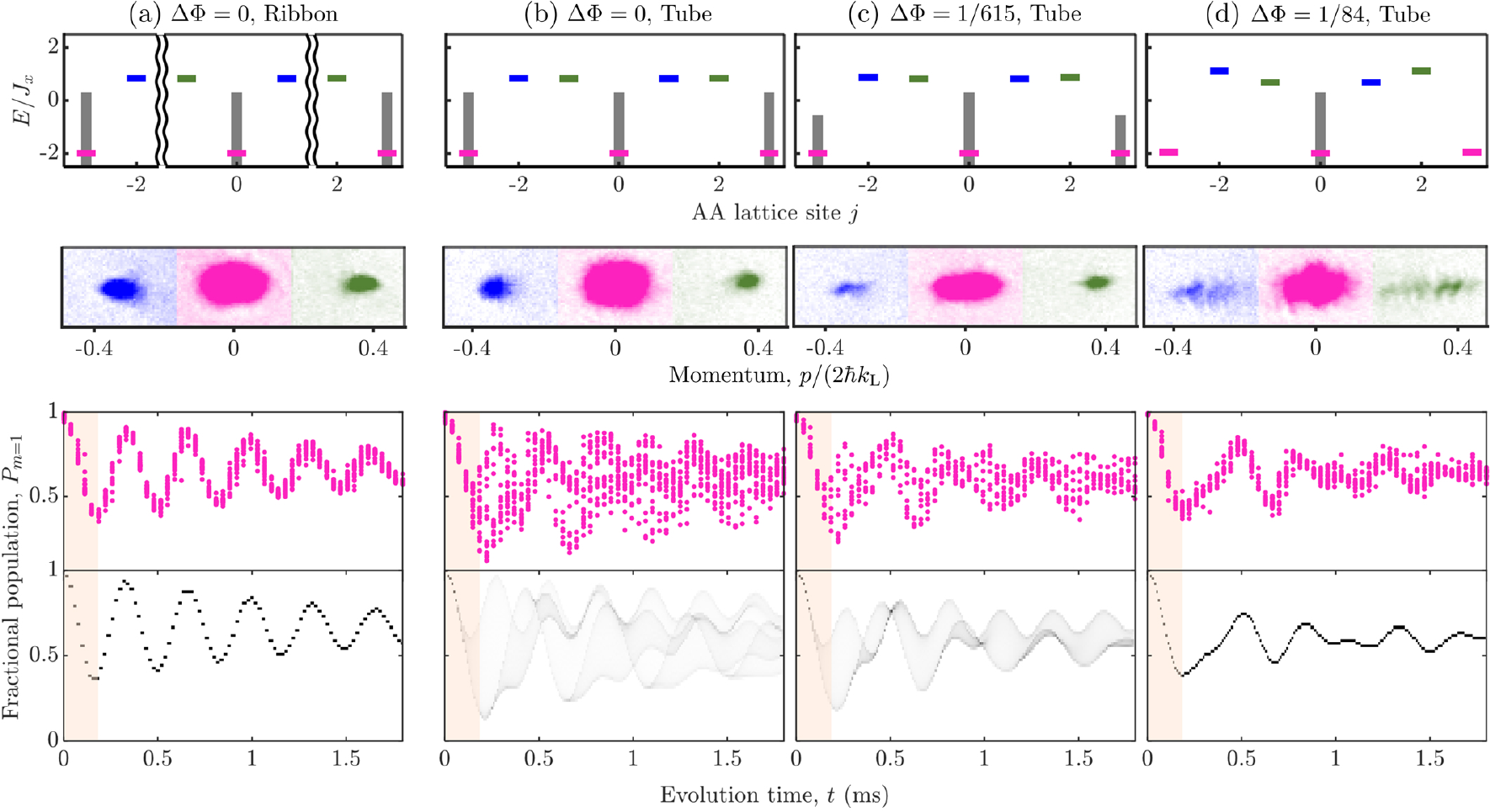}
		\end{centering}
	\caption{{\bf Representative data}
	Each column depicts a value of $\Delta\Phi$, from 0 (showing significant $\phi$-dependence) to $1/84$ (showing negligible $\phi$-dependence), along with a ribbon geometry as a control case.
	The top row depicts representative AA lattices $\ket{j=0}=\ket{m_0=1,q_0=0}$ for each $\Delta\Phi$, along with the initial wavepackets calculated from $R_{\rm TF}=11.5$ $\mu$m (grey bars).  
	The AA lattice schematic for $\Delta\Phi=0$ was calculated with $\Delta\Phi=1/10000$ since the AA ring only has 3 sites at $\Delta\Phi=0$.
	The middle row shows TOF data for each case measured at  $t=1.506\ {\rm ms}$.
	Since diffraction orders outside $q/(2 \hbar k_{\rm L})\in[-1/2,1/2)$ replicate those inside, with amplitude governed by the Wannier orbitals of the longitudinal lattice~\cite{Spielman2007}, we focus on the regime inside.
	The bottom row plots the time evolution of single experimental run data (top) and histogram of the predicted trajectories for all $\phi$ (bottom).
	The shaded regions marks $t<0.185\ {\rm ms}$.
	}
	\label{fig3}
\end{figure*} 
%\twocolumngrid

We began with BECs in $\ket{m=1}$, adiabatically loaded into the ground state of the optical lattice, and initiated dynamics by abruptly introducing $J_{\rm s}$ for a time $t$ up to $1.8\ {\rm ms}$, when the lattice, Raman, and dipole trap lasers were simultaneously extinguished.
During the subsequent $21\ {\rm ms}$ time-of-flight (TOF) we applied a magnetic field gradient to spatially separate atoms in the three $\ket{m}$ states.
We then used absorption imaging to detect the resulting density distribution, yielding the longitudinal momentum distributions of each $\ket{m}$ state.

%%%%%%%%%%%%%%%%%%%%%%
% fig1 (d)(e)
%%%%%%%%%%%%%%%%%%%%%%%%%
% Definition of $\Delta\Phi$
% separation of technical noise and sensitivity
% one experimental run corresponds to one $\phi$
%%%%%%%%%%%%%%%%%%%%%%%%%

%%%%%%%%%%%%%%%%%%%%%%%%%%%%%%%
% introduce AA model
%%%%%%%%%%%%%%%%%%%%%%%%%%%%%%%
% Connect the cylinder to AA model
% Connect the dispersion to the AA model
% Reason for the selection of 2/3
% system parameters in the model
%%%%%%%%%%%%%%%%%%%%%%%%%%%%%%

{\it Model} We quantitatively analyze our experiment by Fourier transforming the HH Hamiltonian along $\ex$ giving $\hat{H}=\sum_{q_0}\hat{H}_{\rm AA}(q_0)$ with
\begin{align}
\hat{H}_{\rm AA}(q_0)=&-2J_{x}\sum_{j}\cos\left[2\pi (j\Phi + \frac{q_0}{ 2 \hbar k_{\rm L}})\right]\ket{j}\bra{j}\nonumber\\
&-J_{\rm s}\left(e^{i\phi}\sum_{j}\ket{j+1}\bra{j}+{\rm h.c.}\right),
\label{eq2}
\end{align}
labeled by crystal momentum $q_0$.
Each $\hat{H}_{\rm AA}(q_0)$ is a realization of the 1D Aubry-Andrey (AA) lattice~\cite{aubry1980analyticity} with nearist-neighbor hopping strength $J_{\rm s}$, sinusoidal potential with depth $4J_{x}$ and phase set by $q_0$. 
The sites of this AA lattice $\ket{j}\equiv\ket{m_0+j,q_0+j\Phi\times 2 \hbar k_{\rm L}}$ are labeled by azimuthal site-index $m$ along with longitudinal crystal momentum $q$.
As shown in Fig.~\ref{fig2}(a), the sinusoidal potential originates from Raman transitions changing the crystal momentum by $2 \hbar k_{\rm L}\Phi$ as $m$ is incremented, in effect sampling the lowest band of the longitudinal lattice.
For rational $\Phi$, each $\hat{H}_{\rm AA}(q_0)$ describes a ring [Fig.~\ref{fig2}(b)] of size $N_{\rm AA}={\rm LCM}(M,Q)$ (${\rm LCM}$ denotes the least common multiple) since $\ket{m_0+N_{\rm AA},q_0+N_{\rm AA}\Phi\times 2 \hbar k_{\rm L}}$ coincides with the initial state $\ket{m_0,q_0}$.
For irrational $\Phi$ (incommensurate potential) the AA model is 1D quasicrystal with a prototypical metal ($J_{\rm s} > J_{x}$) to insulator ($J_{\rm s} < J_{x}$) transition~\cite{Kohmoto1983, qizhou}, and at criticality $J_{\rm s} = J_{x}$ it is a quantum fractal, showing features of quantum chaos~\cite{Evangelou2000}.

\begin{figure*}[t!]
	\begin{centering}
		\includegraphics{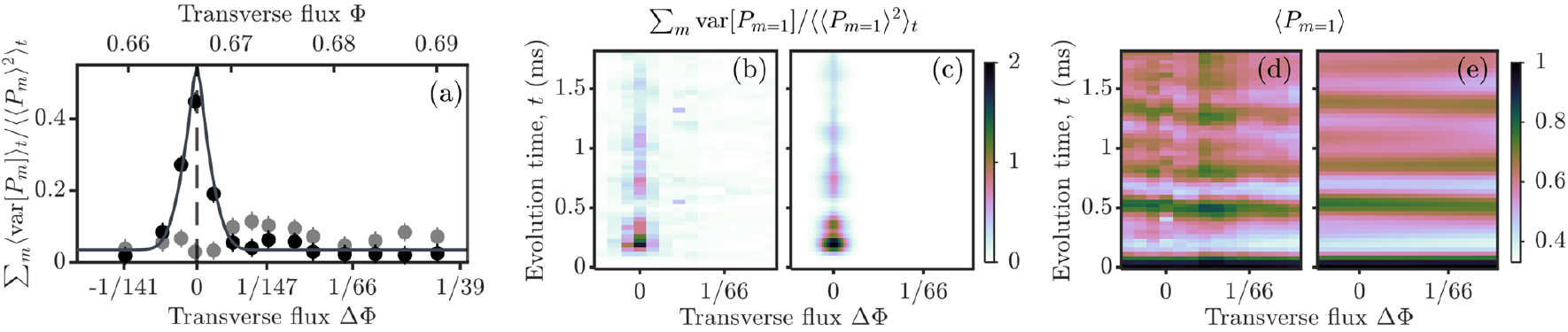}
		\end{centering}
	\caption{\textbf{$\Delta\Phi$ dependance.} ({a}) Phase sensitivity characterized by the time averaged normalized variance is plotted for the fully connected (black) and fragmented (grey) AA rings.
The solid line results from the numerical simulation of the full model with $R_{\rm TF}=11.5\ \mu{\rm m}$, with baseline shifted by the averaged value of the experimental data away from the peak. Normalized variance ({b}) and mean time evolution ({d}) data were linearly interpolated from unequally spaced transverse fluxes probed experimentally. ({c,e}) are their corresponding numerical simulations, respectively.}  
	\label{fig4}
\end{figure*}

%%%%%%%%%%%%%%%%%%%%%%%%%%%%%%%
% wavepackets and interference
%%%%%%%%%%%%%%%%%%%%%%%%%%%%%%%
We focus on the case relevant to our experiment, with $M=3$ and initial site $m = 1$.
Given the BEC's RMS spatial extent $w \approx 0.46 R_{\rm TF} = 20(1)a$ (Appendix~\ref{Appendix: Experimental details}), the corresponding momentum-space density distribution had RMS width $w_q \approx \hbar/(2 w) = 4.0(2)\times 10^{-3}\times 2 \hbar k_{\rm L}$, schematically indicated by the grey Gaussian in the left panel. 
The azimuthal tunneling from Raman coupling induces transitions (tan arrows) that change the crystal momentum by $2 \hbar k_{\rm L}\Phi$, as well as imparting the phase $\phi$. 
Together with $\Phi$, the momentum space width $w_q$ specifies the initial occupation of AA lattice sites: every third AA site potentially samples the initial wavepacket until $3j|\Delta\Phi |\times 2 \hbar k_{\rm L}\gtrsim w_q$ when the site's crystal momentum falls outside the initial wavepacket.
A representative AA lattice is shown in Fig.~\ref{fig2}(b) where the grey bars result from sampling the filled grey curve in Fig.~\ref{fig2}(a).
In the AA model's metallic phase, where our experiments took place, amplitude ballistically expands from each initially occupied site, while in the insulating phase amplitude remains exponentially localized near each initial sites~\cite{qizhou}.
When $3|\Delta\Phi|\times 2 \hbar k_{\rm L}\ll w_q$, the AA ring contains multiple initially occupied sites; as the system evolves, amplitude originating in different sites can overlap and interfere.
In contrast, for $3|\Delta\Phi|\times 2 \hbar k_{\rm L}\gg w_q$, interference is absent due to lack of initial adjacent occupancy.
Any interference depends strongly on $\phi$, which appears in the tunneling term of Eq.~\eqref{eq2}.
We quantify the degree of interference by the normalized variance ${\rm var}[P_m(t)]/\langle P_m(t)\rangle^2$, which is proportional to $\sum_{j\neq0}S(3 \Delta\Phi j\times 2 \hbar k_{\rm L})$, for $3|\Delta\Phi|\times 2 \hbar k_{\rm L}\gtrsim w_q$ [Eq.~\eqref{eq:variance structure factor sum}]. $S(\delta q) \equiv \left|\int dq \psi^*(q+\delta q)\psi(q)\right|^2$ is the static structure factor, and $\psi(q)$ is the momentum-space wavefunction.
%The analytic treatment~\cite{SI} shows that for $3|\Delta\Phi|\times 2 \hbar k_{\rm L}\gtrsim w_q$ this normalized variance is proportional to $S(3 \Delta\Phi\times 2 \hbar k_{\rm L})$, where $f(\delta q) \equiv \int dq \psi^*(q+\delta q)\psi(q)$ is the form factor expressing the overlap integral, and $\psi(q)$ is the momentum-space wavefunction.

{\it Discussion}
Figure~\ref{fig3} summarizes our data for three values of $\Delta\Phi$ and compares the experimental data with our numerical modeling using parameters obtained from fits to the data in Fig.~\ref{fig1}(b). 
These parameters are within the uncertainties (Appendix~\ref{Appendix: simulations}) of independent calibrations.
To achieve quantitative agreement with our data, we include a phenomenological dephasing parameter obtained from fitting the decaying sinusoid of Fig.~\ref{fig3}(a).
The top row depicts the AA lattice along with the initial state (grey bars), showing the transition from multiple sources for $\Delta\Phi \approx 0$, to a single isolated source for $\Delta\Phi = 1/84$. 
In the ribbon case [Fig.~\ref{fig3}(a)], the disconnected links in the lattice, resulting from removing the Raman coupling between $\ket{m=0}$ and $\ket{m=2}$, prevent any potential interference, and the $\phi$-sensitivity vanishes for all transverse flux $\Phi$.

%Figure~\ref{fig2}(c) shows representative TOF data for $\Delta\Phi=1/84$ and initial state $\ket{m=1}$ (middle row).
%In our representative TOF data [Fig.~\ref{fig2}(c)], each crystal momentum state of the longitudinal lattice consists of momentum states that are imaged as horizontally spaced diffraction orders with spacing $2 \hbar k_{\rm L}$. 
%The synthetic lattice sites are resolved vertically; the diffraction orders in these sites are shifted by $2 \hbar k_{\rm L}\Phi$ for each synthetic hopping event as indicated by Fig.~\ref{fig2}(a).
%The elongated diffraction orders in $m=0,2$ are present only for $\Delta\Phi\neq0$, and result from the incommensurate lattice and Raman recoils.
The second row of Fig.~\ref{fig3} plots the three diffraction peaks in the range $-1/2<q/(2 \hbar k_{\rm L})<1/2$, derived from our TOF data [see Fig.~\ref{fig2}(c) for example].
When $\Delta\Phi$ was small enough to allow interference between atoms originating from different AA lattice sites [Fig.~\ref{fig3}(b)(c)], the corresponding crystal momentum difference between different paths to the same order is too small to be resolved. 
When $3|\Delta\Phi|\times 2 \hbar k_{\rm L}\gg w_q$, the momentum difference became resolvable, causing each order in the TOF image to fragment into multiple overlapping sub-orders. 
In the ribbon case, only three orders are present irrespective of $\Delta\Phi$.

The third row of Fig.~\ref{fig3} compares the observed fractional population in the initial state $P_{m=1}$ for individual measurements without averaging (top) with the prediction of our model (bottom).
%Each pink symbol marks the observed fractional population in the initial state $P_{m=1}$ for a single measurement, and the bottom panel is a histogram of the predicted trajectories for all $\phi$.
The experimental data is highly variable only for small $\Delta\Phi$ with a manifestly non-Gaussian distribution.
The noisy dynamics bagin at $t\approx \pi\hbar/(3 J_{\rm s})=0.185\ {\rm ms}$ when population was significantly transferred out of the initial state, and interference became possible~\footnote{For a 1D chain of three sites, population is maximally transferred out of the initial state at $t= \pi\hbar/(3 J_{\rm s})$ with periodic boundaries and $t=\pi\hbar/(2\sqrt{2}J_{\rm s})$ with hard-wall boundaries. }.
The numerical model fully captures the observed spread of data, its time-dependence and even the non-Gaussian distribution, including singular features, i.e., caustics~\cite{Berry1999}.
The ribbon case lacks interference and exhibits only technical noise.

%%%%%%%%%%%%%%%%%%%%%%%%%%%%%%%
% time scales
% 1/(2*np.sqrt(2)*OmegaRaman*Er)
% Out[8]: 0.0001937544269589115
%%%%%%%%%%%%%%%%%%%%%%%%%%%%%%%

%%%%%%%%%%%%%%%%%%%%%%%%%%%%%%%%%
% Describe the absorption images
%%%%%%%%%%%%%%%%%%%%%%%%%%%%%%%%%
% Connect the AA ring to absorption images
% Connect the absorption images to the time evolution
% Explain how to read the absorption images
%%%%%%%%%%%%%%%%%%%%%%%%%%%%%%%%%

The time-averaged normalized variance [Fig.~\ref{fig4}(a)] exhibits
a sharp peak at $\Phi=2/3$ for the tube geometry, despite being featureless for the ribbon geometry.
The measured time dependence of the normalized variance, shown in Fig.~\ref{fig4}(b), is peaked at $\Delta\Phi=0$ for all times, but with variable amplitude. 
In both cases, the numerical calculations [Fig.~\ref{fig4}(c), and solid curve in Fig.~\ref{fig4}(a)] are nearly indistinguishable from the data.
Lastly, the $\phi$-averaged time evolution for both experiment [Fig.~\ref{fig4}(d)] and numerics [Fig.~\ref{fig4}(e)] shows no feature at $\Delta\Phi=0$ where the noise feature is maximal in Fig.~\ref{fig4}(b,c).
This can be directly understood with the HH model which, by a suitable change of $\phi$, transforms site $n$ to $n=0$ [Eq.~\eqref{eq1}].
For irrational transverse fluxes, the system uniformly samples $\phi$, leading to spatially self-averaged time evolution.
Although individual systems at rational flux lack the spatial self-averaging effect, averaging over $\phi$ recovers the uniform sampling in the irrational case, resulting in similar mean time evolution as nearby irrational fluxes (Appendix~\ref{Appendix: AA lattices in real space}).

% Outlook
{\it Outlook} 
The realization of the HH model in the highly-asymmetric geometry proved an ideal testbed of the phenomena in this study, and the choice $\Phi\approx2/3$ minimized the distance between the adjacent initially occupied AA lattice sites for $M=3$.
The onset time for noisy dynamics should increase with $M$, and if $M$ were comparable to the longitudinal extent, we would not have observed any phase sensitivity within the experimental time scale.

Our AA model makes additional predictions beyond the experimental observations presented here.
At longer evolution times than in our experiments, AA rings in the metallic phase with just one initially occupied site can exhibit long-time interference as amplitude fully encircles the ring.  
While this work did not address the question of giving $\phi$ explicit time-dependence, we might expect that the dynamics will be unchanged, provided that both $J_{\rm s}$ and $J_x$ are large compared to the energy shift from the effective electric field $\xi d \phi/dt$, over the extent of the AA localization length $\xi$.
%a simple argument suggests that sufficiently slowly varying $\phi(t)$ will still leave the dynamics unchanged for irrational $\Delta\Phi$ near  $\Phi = 2/3$ and in the AA insulating phase.
%In the momentum space AA picture, $\phi(t)$ is equivalent to the potential $j d \phi/dt$ (i.e., a uniform electric field).
%We might expect that the dynamics will be unchanged, provided that both $J_{\rm s}$ and $J_x$ are large compared to the energy shift from the electric field $\xi d \phi/dt$  over the extent of the  the AA localization length $\xi$.
If true, this would be a new type of quasidisordered-stabilised decoherence-free subspace.

It is worth noting that spatially 1D systems, in which interaction and correlation effects become important---a Hall tube like ours may host exotic topological and magnetic states~\cite{Petrescu2015,Calvanese-Strinati2017,kozarski2018quasi,zhou2020synthetic,Buser2020}. We hope our study will inspire more research in such systems.

\let\oldaddcontentsline\addcontentsline% Store \addcontentsline
\renewcommand{\addcontentsline}[3]{}% Make \addcontentsline a no-op
\begin{acknowledgments}
We thank R.~P.~Anderson for assistance in the early stages of the experiment.
We are grateful to Q.~Zhou for providing Ref.~\onlinecite{qizhou} in advance of submission, E.~Tiesinga and P.~K.~Elgee for careful reading of the manuscript, as well as H.~M.~Hurst, E.~B.~Rozenbaum and W.~D.~Phillips for extended discussions.
Lastly, we appreciate our interactions with M.~Rigol and V.~Galitski who convinced us that this result has absolutely nothing to do with quantum chaos.
This work was partially supported by the AFOSRs Quantum Matter MURI, NIST, and the NSF through the PFC at the JQI. 

\end{acknowledgments}

\appendix

\section{Experimental details}
\label{Appendix: Experimental details}
Each experiment began with a BEC of $\textsuperscript{87}$Rb in a far-detuned crossed dipole trap, with trap frequencies $(f_x,f_y,f_z)\approx(44,53,159)\ {\rm Hz}$ in $\ket{F=1,m_F=-1}$ sublevel, where $F$ and $m_F$ are the total atomic and magnetic angular momentum, respectively.
Owing to the central role of the system size in our discussion, we independently determined the longitudinal Thomas-Fermi radius $R_{\rm TF}$ of the BEC using both TOF and in-situ images. 
We used the Castin-Dum equations~\cite{castin1996bose} to obtain $R_{\rm TF}=11.6(5)\ {\mu}{\rm m}$ from TOF images.
In addition, digitally refocused in-situ images~\cite{Turner2005,Perry2021} gave $R_{\rm TF}=11.4(6)\ {\mu}{\rm m}$.
Because these two measurements share the same magnification, their uncertainties are correlated, and take the average $R_{\rm TF}=11.5(5)\ {\mu}{\rm m}$.
% TF radius from TOF 11.6±0.5
% TF radius from in-situ 11.4±0.6
% along the lattice direction
% 11.5/0.532*2=43.2
%and the Gaussian $e^{-1}$ (atomic distribution $\propto |\psi_0|^2$) width

The rf field for the dynamical decoupling $B_{\rm rf}$ [Fig.~\ref{fig1}(d) in the main text] was assigned a randomized phase just prior to adiabatically loading the atoms into the $\ket{m=0}$ pseudo-spin state.
This mapped the $\ket{m_F=-1,+1,0}$ magnetic sublevels to the $\ket{m=0,1,2}$ pseudo-spin states~\cite{trypogeorgos2018synthetic}, which are insensitive to the magnetic field and facilitate cyclic Raman coupling~\cite{Valdes-Curiel2021}.
The magnetic field $B_0$ and rf strength $\Omega_{\rm rf}$ were stabilized using the procedure detailed in Sec.~\ref{Sec:Magnetic field and rf locks}.
We then applied an rf $\pi$-pulse  to prepare the atoms in the $\ket{m=1}$ state.
This field was linearly polarized the direction orthogonal to both $B_{\rm rf}$ and $B_0$. 
We removed any remaining atoms in $\ket{m=0}$ by transferring them to the $F$=2 hyperfine manifold and blowing them away with a resonant light pulse.
Meanwhile, the atoms were adiabatically loaded into the 1D optical lattice with a ramp duration of $200\ {\rm ms}$.
We then studied dynamics by switched on the Raman coupling for the evolution time $t$, at which point the Raman, optical lattice and dipole trap beams were suddenly extinguished, initiating TOF. 

To achieve the momentum and pseudo-spin resolved imaging, we reversed the process of loading the atoms into the rf dressed states.
Namely, during the $21.3$ ms TOF, we adiabatically ramped down the magnetic field, and then ramped $\Omega_{\rm rf}$ to zero.
As a result, states $\ket{m=0,1,2}$ mapped back to $\ket{m_F=-1,+1,0}$, respectively, which were then separated using the Stern-Gerlach effect with a bias field perpendicular to the lattice beam propagation direction.
The absorption image in Fig.\ref{fig2}(b) in the main text does not show the $\ket{m}$ states in the same relative positions as they appeared on our camera ($\ket{m=1}$ and $\ket{m=2}$ states were switched for logical clarity).
Additionally, due to the high magnetic field gradient applied, a harmonic potential perpendicular to the bias field stretches $\ket{m_F=+1}$ and compresses $\ket{m_F=-1}$ states by $3\sim4\%$.
The optical density profiles of $\ket{m_F=\pm1}$ sublevels were rescaled such that the distances between the neighboring orders, equal to the two-lattice-photon recoil momenta, were the same for all three sublevels.

\subsection{Raman setup}
The Raman beams were almost colinear with the counter-propagating optical lattice beams, with the lattice beam bisecting the angle $\beta=0.34(6)^{\circ}$ between the two Raman beams.
In this configuration, the transverse flux per lattice plaquette was $\Phi=k_R\cos[\beta/2]/k_L\approx k_R/k_L$.
The tiny angle $\beta$ was introduced to avoid retro-reflected beams off of common optics.
The Raman beams were carefully aligned such that there was no momentum transfer along the transverse direction making the dynamics of our system essentially one-dimensional.

We locked the relative phase between between Raman beams, the optical lattice beams and $B_{\rm rf}$.
We detected the beatnote of the two Raman beams in the vicinity of the retro mirror of the optical lattice.
The beatnote had two or three frequency components corresponding to the two or three Raman transitions.
The $\ket{m=0}\leftrightarrow\ket{m=1}$ transition had the largest intensity and was relatively far from the other transitions, and was therefore chosen as the locking frequency.
The local oscillator was taken from the same direct digital synthesized (DDS) signal generator as $B_{\rm rf}$, such that the rf field for the dynamical decoupling and the Raman beatnote were in phase.
We applied feedback to applied to the Raman beam with one frequency component [Raman 1 in Fig.~\ref{fig1}(d) in the main text].
In the rotating frame, the energy differences between the two Raman beams were $\delta\epsilon_{m}\equiv \epsilon_{m}-\epsilon_{m+1}$, where $m=0,1,2$.
In the lab frame, they are $\delta\epsilon_{m}-\epsilon_{\rm rf}$, so the beatnote frequencies were $\approx22$ MHz~\cite{anderson2020realization}. 

\subsection{Magnetic field and rf locks}\label{Sec:Magnetic field and rf locks}

We then applied two microwave pulses of duration $T_{\mu w}=100$ $\mu$s, separated by 1/60 s to partially transfer $\sim5\%$ of the atoms to $\ket{F=2,m_F=-1}$ sublevel.
At the locking point, the two pulses were blue and red detuned by $1/(2T_{\mu w})$ from the resonance.
The lock point $B_{\rm lock}\approx3.137$ mT gave an rf frequency $\omega_{\rm rf}/(2\pi)=22.1$ MHz resonant with the $\ket{F=1,m_F=-1}\leftrightarrow\ket{F=1,m_F=0}$ transition.
The error signal was the imbalance of the two transfers
\begin{align*}
\epsilon &=\frac{{\rm OD_{int}^{(1)}-OD_{int}^{(2)}}}{{OD_{int}^{(1)}+OD_{int}^{(2)}}},
\end{align*}
where ${\rm OD_{int}^{(i)}}$, $i=1,2$, was the integrated optical density of the two images.
This error signal drove a slow proportional-integral-differential (PID) lock that set the magnetic fields in subsequent reputations of the experiment.
Our procedure is extends earlier work~\cite{leblanc2013direct}, and for further specifics see Ref.~\onlinecite{si_Anapaper}.

After the atoms were transferred to $\ket{m=0}$ state, we used a similar technique to stabilize the strength of the rf field $\Omega_{\rm rf}$.
The microwave pulses drove the transition with the largest sensitivity to $\Omega_{\rm rf}$ and the least sensitivity to the $B_0$, namely, from $\ket{m=0}$ to the state that asymptotically goes to $\ket{F=2,m_F=-2}$ with decreasing magnetic field.
  
\section{Reduction to HH model}
We start with the full Hamiltonian
\begin{align}
\hat H_{\rm full} =& \left[\frac{\hbar^2 \hat k^2}{2 M_a} + \frac{V_{\rm L}}{2}\cos(2\kl \hat x) + V_{\rm ext}(\hat x)\right]\otimes \hat 1 + \nonumber\\
&\left[\sum_m \delta_m \ket{m}\bra{m} \right] - \label{ap:eq:full_Hamiltonian}\\
& \left[\sum_m \frac{\Omega_{{\rm R},m}}{2}e^{i (2 \kr \hat x+\phi)}\ket{m+1}\bra{m} + {\rm H.c.}\right]\nonumber,
\end{align}
describing the light-matter interaction of our three-state atoms of mass $M_a$ including a state independent confining potential $V_{\rm ext}(\hat x)$.
In our experiment, we measured the $V=5.0(1) E_{\rm L}$ lattice depth by suddenly applying the lattice potential and fitting the resulting Kapitza-Dirac time evolution~\cite{Kapitza1933}.
We obtained the Raman coupling strength $\Omega_{{\rm R},m} = \Omega_{\rm R} = 0.296(6)E_L$ by separately measuring the Rabi frequencies of each transition and adjusting them to be equal within our uncertainties.
This process was repeated each time we changed the wavelength of the titanium sapphire laser producing the Raman laser beams.
The parameters $\delta_m = 0.00(2)E_{\rm L}$ describe detuning from Raman resonance.
The stated uncertainties include the variation across the measured transverse fluxes.

The ground-band behavior of $\hat H_{\rm full}$ can be approximated by the tight binding HH Hamiltonian
\begin{align}
\hat{H}(\phibar)=&-J_x\sum_{m,n}\ket{m,n+1}\bra{m,n}-\nonumber\\
&J_{\rm s}\sum_{m,n}e^{i2\pi(\Phi n+\phibar)}\ket{m+1,n}\bra{m,n}+{\rm h.c.},
\label{eq: HH Hamiltonian}
\end{align}
where in analogy to the relation between the Planck constant $h$ and $\hbar\equiv h/(2\pi)$, we define $\phibar\equiv\phi/(2\pi)$.
We connected these two models by numerically solving the first term in Eq.~\eqref{ap:eq:full_Hamiltonian} to obtain the ground-band Wannier orbitals $w_0(x-n a)$ for each lattice site $n$, from which we obtained the longitudinal tunneling strength
\begin{align*}
J_x &= -\int dx \,w^*_0(a)\left[-\frac{\hbar^2}{2M_a}\frac{d^2}{dx^2}+\frac{V_{\rm L}}{2}\cos(2\kl \hat x)\right]w_0(0)\\
&= 0.066(2) E_{\rm L}.
\end{align*}
We then projected the second term in Eq.~\eqref{ap:eq:full_Hamiltonian} to the lowest band subspace giving the synthetic dimension tunneling strength
\begin{align*}
J_{\rm s} &= \frac{\Omega_{{\rm R}}}{2} \int dx |w_0(x)|^2 e^{2 i \kr x}= 0.111(3)E_{\rm L}.
\end{align*}
where the integral defines a Lamb-Dicke suppression factor, equal to $0.75$ for our lattice depth.

\section{Simulations}
\label{Appendix: simulations}
\begin{figure}[t]
\begin{center}
\includegraphics{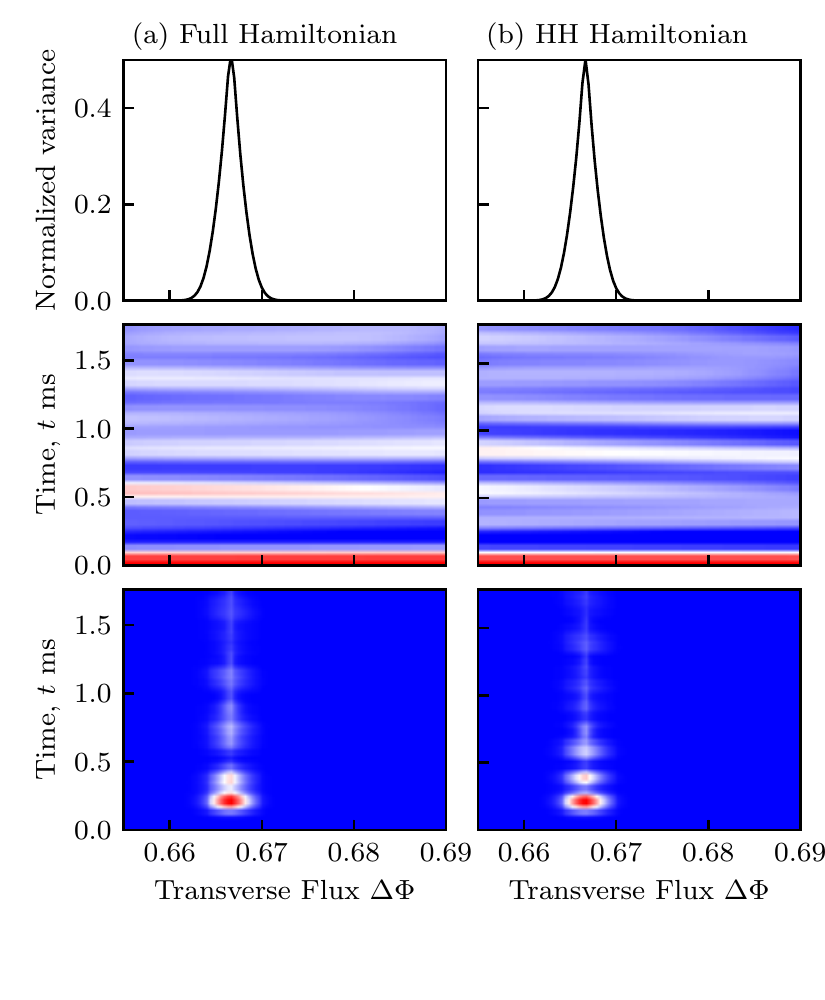}
\end{center}
\caption{\textbf{Full and reduced model comparison} 
(a) Simulation using the full light matter Hamiltonian, and (b) Simulation using the reduced HH Hamiltonian.
Both cases modeled a $1.8\ {\rm ms}$ evolution time.
The first row plots the normalized variance averaged over the evolution time as a function of $\Phi$
The second row plots the occupation probability $P_{m=1}(t)$ of the initial state, as a function of both $\Phi$ and $t$.
The third row shows the normalized variance, as a function of both $\Phi$ and $t$.
}
\label{app:fig:simulations}
\end{figure} 

All of the numerical simulations presented in the main manuscript were complete real space 1D discrete variable representation (DVR) simulations of the full light-matter Hamiltonian~\cite{Colbert1992}.
Here we compare full-system simulations to those of the reduced Hofstadter model, and show: (1) their qualitative time-evolution is the same, but after long times their behavior differs quantitatively; and (2) the predicted peaks in normalized variance as a function of $\Phi$ are indistinguishable.
We used a real space DVR method rather than momentum space band structure approach for two reasons: (1) band structure simulations are not possible for irrational flux, since there is no periodic potential; and (2) our observed noise signature depended critically on the spatial extent of our system.

Our simulations of non-interacting atoms included a harmonic potential $V_{\rm ext}$ that served to define the spatial extent of the ground state wavefunction
\begin{align*}
\psi_{\rm HO}(x) &\propto \exp\left[-\frac{1}{2}\left(\frac{x}{\ell_{\rm HO}}\right)^2\right],
\end{align*}
with harmonic oscillator length $\ell_{\rm HO}$.
Because the static structure factor $S(q)$ governs the coherence peak width, we selected the frequency of the harmonic potential frequency so the RMS width of $S(q)$ from the numerical simulation was equal to that resulting from  $n(x) \propto [1-(x/R_{\rm TF})^2]^2$, the 1D profile derived by integrating a 3D Thomas-Fermi profile along both transverse directions.
This yields the harmonic oscillator length $\ell_{\rm HO} = 20 \sqrt{2\pi} R_{\rm TF}/77 \approx 0.651 R_{\rm TF}$ giving a spatial density profile with RMS width $w = \ell_{\rm HO}/\sqrt{2} \approx 0.460 R_{\rm TF}$.
The associated widths in momentum space are
$\kappa_{\rm HO} = 1/\ell_{\rm HO}$ with a momentum-density RMS width $w_q = 1/(2 w)$.
With this definition of the potential, every term in both the light matter Hamiltonian and the HH Hamiltonian are fully defined.

We validated our calibrations by performing least squares fits to the data in the middle panel of Fig.~\ref{fig1}(b), including $P_0(t)$ and $P_2(t)$, for a ring-coupling geometry at $\Delta\Phi = -1/141$, as well as the data in Fig.~\ref{fig3}(d) for the ribbon geometry.  
The resulting best fit coefficients $(\Omega_{{\rm R},0}, \Omega_{{\rm R},1}, \Omega_{{\rm R},2}) = (1.04, 0.98, 0.98)\Omega_{\rm R}$ and $(\delta_0, \delta_1, \delta_2) = (0,0,0.02) E_{\rm L}$ are consistent with the uncertainties of our independent calibrations.
All of our simulations include a single phenomenological fitting parameter $\tau=1.5(2)\ {\rm ms}$, obtained only from the fit to the ribbon geometry data, to capture the slow decay of the observed coherent evolution.

We conclude with a side-by-side comparison of the dynamics of these two descriptions, as depicted in Fig.~\ref{app:fig:simulations}, with the full light matter simulation in (a) and the HH description in (b).
Firstly, the time-averaged noise variances, plotted in the first row as a function of $\Phi$, are indistinguishable confirming that our key observation is a property both of our true physical system as well as the reduced HH model.
The second and third rows display the time evolution of the initial state probability $P_{m=1}(t)$ and the noise variance, respectively.
These data show that the time evolution of these simulations share the same dynamical time scale and qualitative features, but differ markedly in their quantitative evolution.
Still in both cases, the probability $P_{m=1}(t)$ shows no feature associated with the peak in the noise variance.

\section{AA lattices in momentum space}

To analytically describe the variance of the time evolution in our experiments, we reduce the 2D HH model to a collection of independent 1D AA models.
We begin by Fourier transforming the real dimension 
\begin{equation}
\ket{m,n}=\int_{-1/2}^{1/2} d q\ e^{-i2\pi q n}\ket{m,q},
\end{equation} 
to realize a series of AA lattices in momentum space: 
\begin{align*}
\HAA(\qi)=&-2J_x\sum_{j}\cos[2\pi (\qi+j\Phi)]\ket{j;\qi}\bra{j;\qi}\\
&-J_s\left(e^{i\phi}\sum_{j}\ket{j+1;\qi}\bra{j;\qi}+{\rm h.c.}\right).
\end{align*}
In the main text, we approximated irrational transverse fluxes by nearby rational numbers within the finite resolution of the system.
Here, we instead approximate rational numbers by neary irrational numbers.
In this case, each AA Hamiltonian $\HAA(\qi)$ describes an infinite chain, and the total Hamiltonian $\hat{H}$ sums over an infinite set of such AA chains, where the crystal momentum $\qi$ is written in units of two-photon recoil momenta $2\hbar k_{\rm L}$.
Here, both the pseudo-spin and crystal momentum are labeled periodically, namely $\ket{m+M}=\ket{m}$ and $\ket{q+1}=\ket{q}$.
Each chain contains a countably infinite set of sites labeled by $\ket{j;\qi}\equiv\ket{m,q}=\ket{\mi+j,\qi+j\Phi}$, where $j\in\mathbb{Z}$. In what follows, we take $\mi = 0$ without any loss of generality.

\begin{figure*}[t]
\begin{centering}
\includegraphics{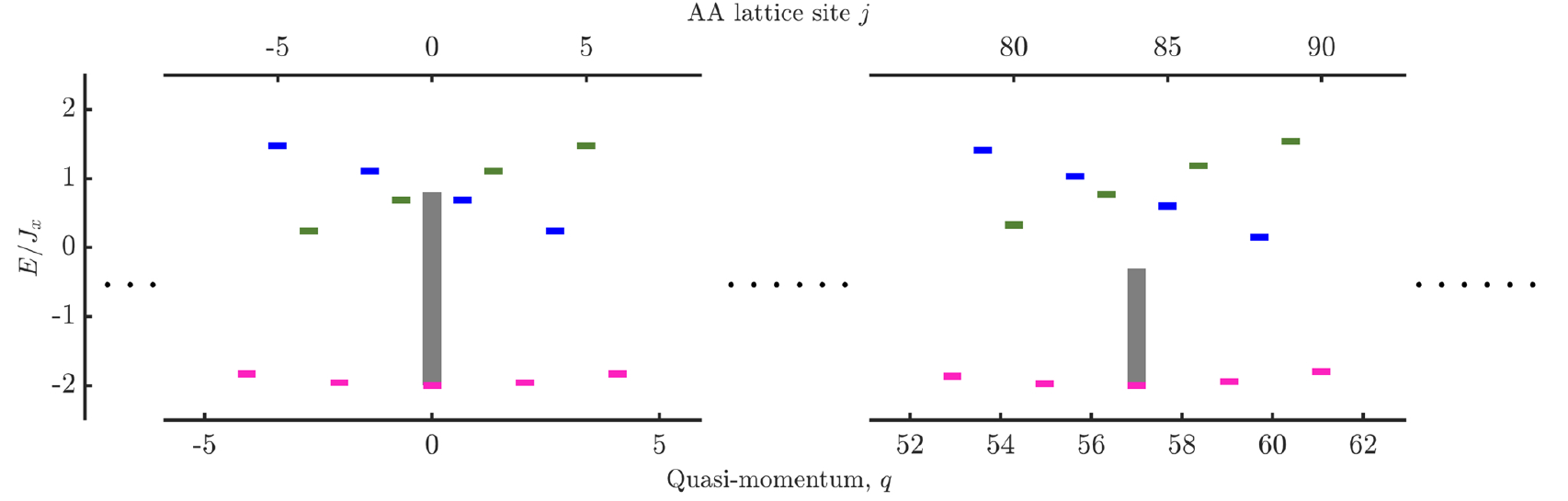}
	\end{centering}
\caption{\textbf{Irrational flux AA chains} An AA chain with $\qi=0$, $\Phi=\P/\Q+\Delta\Phi$ where $\Delta\Phi=1/84+(\sqrt{5}-1)/15000$ and $\P/\Q=2/3$.
The grey bars show the unraveling of a Gaussian wavepacket with width $w_q=1/(2\pi w)$, where $w=23$.}
\label{fig:AA chain}
\end{figure*}

Figure~\ref{fig:AA chain} shows an example of an AA chain with $\Phi = \P/\Q + \Delta\Phi$, and $\Delta\Phi = 1/84 + (\sqrt{5}-1)/15000$, close to the simple rational fraction $\P/\Q = 2/3$.
This chain is nearly indistinguishable from the corresponding AA ring, with $N_{\rm AA}={\rm LCM}(M,Q)=84$ sites, where the sites around $j=84$ are near-replicas of those around $j=0$.
Thus, encircling a rational-flux-ring of size $N_{\rm AA}$  corresponds to traveling between sites of an irrational-flux-chain spaced by $N_{\rm AA}$ sites.
We note that even AA rings can contain these near-replicas and they are spaced by $N_{\rm rep} =  M/[{\rm LCM}(M, \Q) |\Delta\Phi|]$, provided $|\Delta\Phi|\ll{\rm LCM}(M, \Q)|\Delta\Phi|\lesssim w_q$.
At long times, beyond those probed in our experiments, we expect to see additional growth of the variance as the atoms travel to the nearest replica, even for the $\phi$-sensitivity suppressed cases at short time scales.

\subsection{Mean time evolution}
The initial state is a wavepacket of width $w_q$ in momentum space with initial pseudo-spin state $\ket{\mi=0}$:
\begin{align*}
\ket{\tilde{\psi}} &= \int d\qi\tilde{\psi}_0(\qi)\ket{0,\qi}.
%\label{eq:unraval into momentum states}
\end{align*}
This is unraveled in the AA lattices as
\begin{align}
\ket{\tilde{\psi}}&=\int d\qi D(\qi)\sum_j \tilde{\psi}^{\rm AA}_{0,Mj}(\qi) \ket{Mj;\qi},
\label{eq:total wavefunction unravelling}
\end{align}  
where $\tilde{\psi}^{\rm AA}_{0,j}(\qi)=\tilde{\psi}_0(\qi+j\Phi)$ denotes the wavefunction in the chain containing site $\ket{j=0;\qi}=\ket{0,\qi}$. The sum over $j$ includes all sites within a chain. $D(\qi)$ is an everywhere discontinuous function and takes the value either 1 or 0, reminiscent of the Dirichlet function, such that the integral over $\qi$ (summing over infinite chains) includes all the momenta within $[0,1)$ without repeating the momenta already connected within a chain. 

At time $t$, an initial site $\ket{j;\qi}$ evolves to
\begin{align*}
|j(t);\qi\rangle&=\sum_{j'}\alpha_{j,j'}(t;\qi)e^{i\phi(j'-j)}|j';\qi\rangle
\end{align*}
where $\alpha_{j,j'}(t;\qi)=\bra{j';\qi}\UAA(t;\qi)\ket{j;\qi}$ describes the time evolution from site $j$ to $j'$ with $\phi=0$.
We can therefore interpret $\alpha_{j,j'}(t;\qi)$ as the wavefunction at time $t$ for a particle starting in $|j^\prime;\qi\rangle$ expressed in the $|j;\qi\rangle$ basis, with $\phi=0$.

When $\phi\neq0$, the time evolution only contributes a phase factor $\phi(j'-j)$, as can be seen from the expansion of the evolution operator:
\begin{align*} 
\UAA&=e^{-\frac{i\HAA t}{\hbar}}=\mathbb{I}-\frac{it\HAA}{\hbar}+\frac{1}{2!}\left(-\frac{it\HAA}{\hbar}\right)^2+\cdots
%\label{eq:evolution operator expansion}
\end{align*}
The initial state has pseudo-spin $Mj_1$, $j_1\in\mathbb{Z}$. Consequently, the total wavefunction of chain $\qi$ at time $t$ evolves to
\begin{align*}
\ket{\tilde{\psi}(t);\qi}&= \sum_{j_1, j_2} \tilde{\psi}^{\rm AA}_{0,Mj_1}(\qi) \alpha_{Mj_1,j_2}(t;\qi) e^{i\phi (j_2-Mj_1)}\ket{j_2;\qi}
\end{align*}
with the probability to arrive in the final site $\ket{j;\qi}$ equal to
\begin{widetext}
\begin{align}
&\tilde{P}_j(t;\qi)
= \left|\left\langle j;\qi\middle|\tilde{\psi}(t);\qi\right\rangle\right|^2
=\sum_{j_1, j_2} \tilde{\psi}^{{\rm AA}*}_{0,Mj_1}(\qi)\tilde{\psi}^{{\rm AA}}_{0,Mj_2}(\qi) \alpha^*_{Mj_1,j}(t;\qi)\alpha_{Mj_2,j}(t;\qi) e^{i\phi M(j_1-j_2)}
\label{eq: probability to site j}
\end{align}
\end{widetext}
Inserting $\left<\exp[i\phi M(j_1-j_2)]\right>_\phi = \delta_{j_1, j_2}$ into the above equation, leads to the averaged probabilities:
\begin{align*}
\left<\tilde{P}_j(t;\qi)\right>_\phi &= \sum_{j_1} \left|\tilde{\psi}^{\rm AA}_{0,Mj_1}(\qi)\right|^2 \left| \alpha_{Mj_1,j}(t;\qi)\right|^2
\end{align*}
This is an incoherent sum over the probabilities of the atoms starting from each initial site weighted by the probability of traveling from that site to the final site. 

Assuming the initial wavepacket is very narrow, the Hamiltonian in the vicinity of an initially occupied site $\ket{j_1;\qi}$ is close to that at $\ket{j_1=0;\qi}$ and independent of $\qi$. This allows us to replace the evolution from site $j_1$ to site $j$ with that from site 0 to $j-j_1$, i.e., $\alpha_{j_1,j}(t;\qi)\rightarrow\alpha_{0,j-j_1}(t)$, leading to probability depending only on the distance traveled. 

Our experiments measured the probability
\begin{align*}
\left< P_{\mf}(t)\right>_\phi &= \int d\qi D(\qi) \sum_\ell  \langle \tilde{P}_{\mf+M\ell}(t;\qi)\rangle_\phi\\
%\approx   \int d\qi D(\qi) \sum_{\ell_1, \ell_2} \left|\tilde{\psi}^{\rm AA}_{0,M \ell_1}(\qi)\right|^2 \left| \alpha_{0, \mf+ M( \ell_2 - \ell_1)}(t)\right|^2\\
%&= \int d\qi D(\qi)  \sum_{\ell_1} \left|\tilde{\psi}^{\rm AA}_{0,M \ell_1}(\qi)\right|^2 \sum_{\ell_2} \left| \alpha_{0,\mf+M\ell_2}(t)\right|^2
&\approx \sum_{\ell} \left| \alpha_{0,\mf+M\ell}(t)\right|^2
%\label{eq:ensemble_average}
\end{align*}
to arrive at the final state with pseudo-spin $\mf+M\ell$, $\ell\in\mathbb{Z}$, where we made the approximation $\alpha_{j_1,j}(t;\qi)\rightarrow\alpha_{0,j-j_1}(t)$ and reversed the unraveling in Eq.~\eqref{eq:total wavefunction unravelling}.
This equation depends only on the evolution of a particle starting at AA lattice site $\ket{j=0}$. We consider a small range of transverse fluxes in the vicinity of a rational fraction $\Phir$ of the flux quantum. Within the range of propagation, the small deviation of the transverse flux is not resolved, i.e. $\alpha_{0,j-j_1}^\Phi(t)\approx\alpha_{0,j-j_1}^{\Phir}(t)$. We hereby conclude that the mean time evolution in the vicinity of $\Phir$ is a smooth function of $\Phi$. 

\subsection{Variance of the time evolution}
The time evolving variance for arriving in site $m_f$ over a uniformly sampled ensemble of Peierls phase $\phi$ is 
\begin{equation}
{\rm var}_{\mf}(t)\equiv \langle P_{\mf}(t)^2\rangle_\phi-\langle P_{\mf}(t)\rangle^2_\phi
\label{eq: Var def q}.
\end{equation} 
We calculate ${\rm var}_{\mf}(t)$ beginning with 
\begin{widetext}
\begin{equation}
\begin{aligned}
\left< P^2_{\mf}(t)\right>_\phi 
=& \int dq_1D(q_1) \int dq_2D(q_2)\sum_{\ell_1,\ell_2}  \left< \tilde{P}_{\mf+M\ell_1}(t; q_1) \tilde{P}_{\mf+M\ell_2}(t; q_2) \right>_\phi
%\nonumber\\
%\approx& \int dq_1D(q_1) \int dq_2D(q_2)\sum_{j_1, j_1^\prime} \sum_{j_2, j_2^\prime} \left<e^{i M\phi(j_1+j_2-j_1^\prime-j_2^\prime)} \right>_\phi
%\left[ \tilde{\psi}^{{\rm AA}*}_{0,Mj_1}(q_1) \tilde{\psi}^{{\rm AA}}_{0,Mj_1^\prime}(q_1) \tilde{\psi}^{{\rm AA}*}_{0,Mj_2}(q_2) \tilde{\psi}^{{\rm AA}}_{0,Mj_2^\prime}(q_2)  \right]
%\nonumber\\
%&\times \sum_{\ell_1,\ell_2}\left[ \alpha^*_{0,\mf+M(\ell_1-j_1)}(t)\alpha_{0,\mf+M(\ell_1-j_1^\prime)}(t)\alpha^*_{0,\mf+M(\ell_2-j_2)}(t)\alpha_{0,\mf+M(\ell_2-j_2^\prime)}(t)\right] 
%\nonumber\\
%=& \int dq_1D(q_1) \int dq_2D(q_2)\sum_{j_1, j_1^\prime} \sum_{j_2, j_2^\prime}
%\left<e^{i M\phi(j_1+j_2)} \right>_\phi 
%\left[ 
%\tilde{\psi}^{{\rm AA}*}_{0,M(j_1+j_1')}(q_1) 
%\tilde{\psi}^{{\rm AA}}_{0,Mj_1^\prime}(q_1) 
%\tilde{\psi}^{{\rm AA}*}_{0,M(j_2+j_2')}(q_2)
%\tilde{\psi}^{{\rm AA}}_{0,Mj_2^\prime}(q_2) 
%\right]
\\
%&\times \sum_{\ell_1,\ell_2}\left[ \alpha^*_{0,\mf+M(\ell_1-j_1)}(t)\alpha_{0,\mf+M\ell_1}(t)\alpha^*_{0,\mf+M(\ell_2-j_2)}(t)\alpha_{0,\mf+M\ell_2}(t)\right] 
%\\
\approx & \sum_{j}\left[\int dq_1D(q_1) \int dq_2D(q_2) \sum_{j_1', j_2^\prime}
\tilde{\psi}^{{\rm AA}*}_{0,M(j+j_1')}(q_1) 
\tilde{\psi}^{{\rm AA}}_{0,Mj_1^\prime}(q_1) 
\tilde{\psi}^{{\rm AA}*}_{0,M(-j+j_2')}(q_2)
\tilde{\psi}^{{\rm AA}}_{0,Mj_2^\prime}(q_2) 
\right]
\\
&\times \sum_{\ell_1,\ell_2}\left[ \alpha^*_{0,\mf+M(\ell_1-j)}(t)\alpha_{0,\mf+M\ell_1}(t)\alpha^*_{0,\mf+M(\ell_2+j)}(t)\alpha_{0,\mf+M\ell_2}(t)\right] 
\label{eq: p squared average}
\end{aligned}
\end{equation}
\end{widetext}
%where we have changed the variables as $j_i-j_i'\rightarrow j_i$ and $\ell_i-j_i'\rightarrow\ell_i$, $i=1,2$ in the second equality and 
where we have used Eq.~\eqref{eq: probability to site j} and $\left<e^{i M\phi(j_1+j_2)} \right>_\phi=\delta_{j_1,-j_2}$. The expression in the square bracket in second line is 
\begin{equation}
\begin{aligned}
%=&\int dq_1D(q_1) \int dq_2D(q_2) \sum_{j_1', j_2^\prime}
%\tilde{\psi}^{*}_0\left[q_1+M(j+j_1')\Phi\right]
%\tilde{\psi}_0\left[q_1+Mj_1'\Phi\right] 
%\tilde{\psi}^{*}_0\left[q_2+M(-j+j_2')\Phi\right]
%\tilde{\psi}_0\left[q_2+Mj_2'\Phi\right] 
%\\ 
%=& \int dq_1 dq_2 \tilde{\psi}^{*}_0(q_1+M\Phi j)\tilde{\psi}_0(q_1) \tilde{\psi}^{*}_0(q_2-M\Phi j) \tilde{\psi}_0(q_2) 
[\cdot]&=\left|\int dq_1 \tilde{\psi}^{*}_0(q_1+M\Phi j)\tilde{\psi}_0(q_1) \right|^2 
\equiv S(M\Phi j),
%&= |f(M\Phi j)|^2
\label{eq:wavefunc overlap}
\end{aligned}
\end{equation}
where $S$ is the static structure factor~\cite{marder2010condensed} and the normalization implies $S(0)=1$.
We arrived at this expression by reversed the unraveling of the momentum space wavefunction (noted below Eq.~\eqref{eq:total wavefunction unravelling}), i.e. $\int dq_iD(q_i)\sum_{j_i'}\rightarrow\int dq_i$, $i=1,2$.

Combining Eqs.~\mref{eq: p squared average,eq:wavefunc overlap} we arrive at
\begin{equation}
\begin{aligned}
&\left< P^2_{\mf}(t)\right>_\phi 
\approx\sum_j S(M\Phi j)
\\
\times & \left|\sum_\ell\alpha^*_{0,\mf+M(\ell-j)}(t)\alpha_{0,\mf+M\ell}(t)\right|^2.
\end{aligned}
\end{equation}
Importantly, if we also perform a time average we de-phase the cross-terms in the double sum
\begin{align*}
& \left<\cdots\right>_t 
% &= \sum_{\ell_1,\ell_2} \left<\alpha^*_{0,\mf+M[\ell_1-j]}(t) \alpha_{0,\mf+M\ell_1}(t) \alpha_{0,\mf+M[\ell_2-j]}(t) \alpha^*_{0,\mf+M\ell_2}(t)\right>_t \\
% &= \sum_{\ell_1,\ell_2} \left<\alpha^*_{0,\mf+M[\ell_1-j]}(t) \alpha_{0,\mf+M\ell_1}(t) \alpha_{0,\mf+M[\ell_2-j]}(t) \alpha^*_{0,\mf+M\ell_2}(t)\right>_t (\delta_{\ell_1, \ell_2} + \delta_{j,0} - \delta_{j,0}\delta_{\ell_1, \ell_2} ) \\
 = \delta_{j,0} \left<\left[\sum_{\ell} \left|\alpha_{0,\mf+M\ell}(t)\right|^2\right]^2\right>_t 
\\ 
 +& (1- \delta_{j,0})\sum_{\ell} \left<|\alpha_{0,\mf+M(\ell-j)}(t)|^2|\alpha_{0,\mf+M\ell}(t)|^2\right>_t
\end{align*}
The Kronecker delta functions locate those cases that match conjugated terms with non-conjugated terms, and the last term avoids double counting.  

We now shift the first term in the last line to the LHS to get the time-averaged variance
\begin{widetext}
\begin{align*}
\langle{\rm var}_{\mf}\rangle_t &= \sum_{j\neq0} S(M \Phi j) \sum_{\ell} \left<|\alpha_{0,\mf+M(\ell-j)}(t)|^2|\alpha_{0,\mf+M\ell}(t)|^2\right>_t\nonumber\\
\langle{\rm var}\rangle_t &= \sum_{j\neq0} S(M \Phi j) \sum_{i} \left<|\alpha_{0,i-M j}(t)|^2|\alpha_{0,i}(t)|^2\right>_t,
\end{align*}
\end{widetext}
having used $S(0) = 1$. 
In the second line, we defined the total variance by summing over final states. 

\begin{figure*}
\begin{center}
\includegraphics{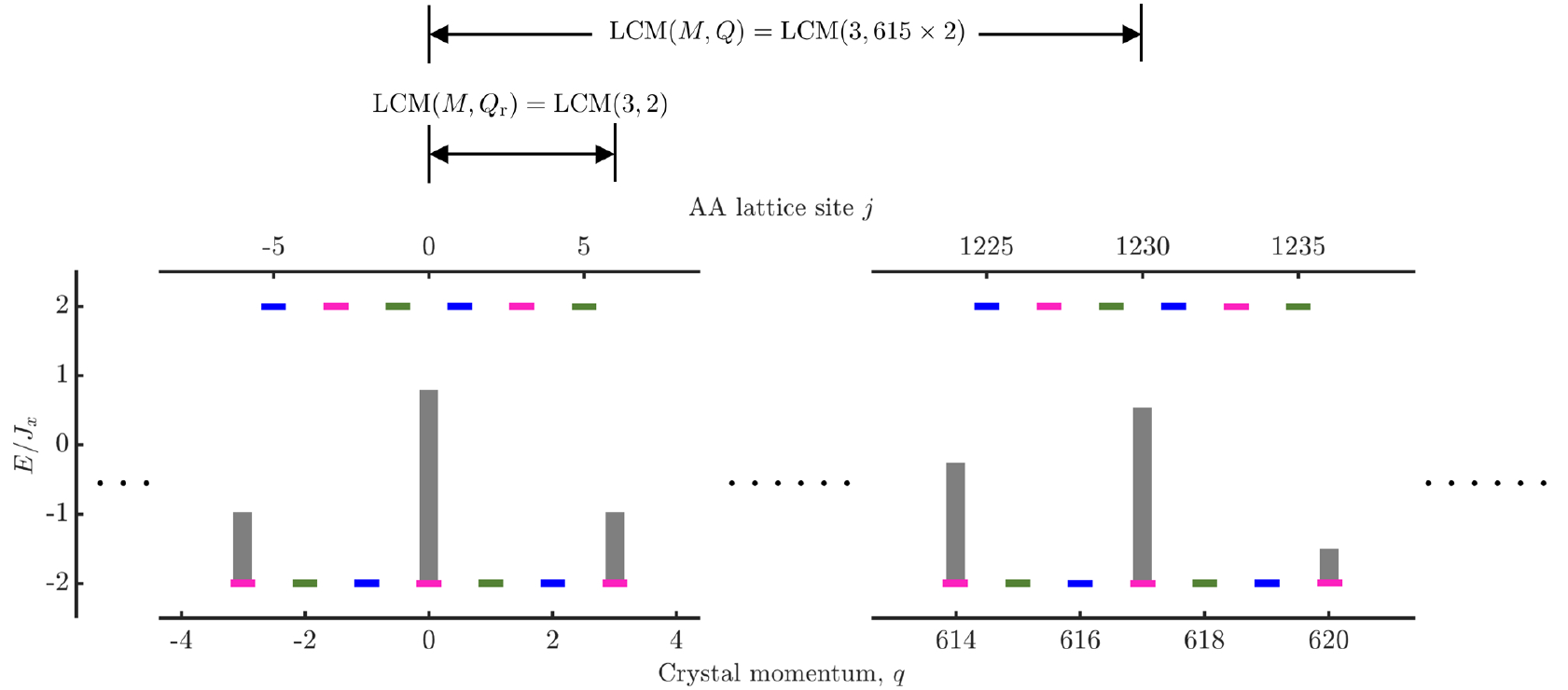}
\end{center}
\caption{\textbf{Irrational flux AA chains with $M\neq{\rm LCM}(M,\Q)$.} An AA chain with $\qi=0$, $\Phi=\P/\Q+\Delta\Phi$ where $\Delta\Phi=1/615+(\sqrt{5}-1)/500000$ and $\P/\Q=1/2$.
The grey bars show the unraveling of a Gaussian wavepacket with width $w_q=1/(2\pi w)$, where $w=23$. These parameters were not probed in our experiments.}
\label{fig:gallary of fluxes}
\end{figure*}

Due to the narrow wavepacket assumption, the sites that could possibly contribute to the structure factor $S$ are separated by ${\rm LCM}(M,\Q)$, rather than $M$ (Fig.~\ref{fig:gallary of fluxes}), where the nearby rational transverse flux $\Phir=\P/\Q$ with $\P$ and $\Q$ being co-prime. In the main text and the experiments, $\P/\Q=2/3$, while here we derive a general case. Therefore, 
\begin{align*}
&S(M\Phi j) =  S[M\Phi j] \delta_{Mj, {\rm LCM}(M,\Q) j} \\
= &S\left[\Delta\Phi_{M,\Q} j\right] \delta_{Mj, {\rm LCM}(M,\Q) j}
\end{align*}
Where we introduced $\Delta\Phi_{M,\Q}\equiv \Delta\Phi {\rm LCM}(M,\Q)$.
Only the non-integer part of the variable matters because the crystal momentum $q$ is periodically labeled. If $|\Delta\Phi_{M,\Q}|\gg w_q$, then the structure factor and hence the variance approach zero for $j\neq0$.
When $|\Delta\Phi_{M,\Q}|\lesssim w_q$, the closest distance between initially occupied sites on the chain is ${\rm LCM}(M,\Q)$ (Fig.~\ref{fig:gallary of fluxes}).
This leads to the final form of the total variance
\begin{align*}
\langle{\rm var}\rangle_t =& \sum_{j\neq0} S[\Delta\Phi_{M,\Q} j]
\\
& \times \sum_{i} \left<|\alpha_{0,i-{\rm LCM}(M,\Q) j}(t)|^2|\alpha_{0,i}(t)|^2\right>_t,
\end{align*}
We now consider the limiting behavior of this function. 

(1) For $|\Delta\Phi_{M,\Q}|\gg w_q$, only $S(0)$ is practically nonzero and we get $\langle{\rm var}\rangle_t = 0$.

(2) For $|\Delta\Phi_{M,\Q}|\gtrsim w_q$, assume that enough time has passed so the wavepackets have spread over a range wide compared to the initial distribution, giving 
\begin{align}
\langle{\rm var}\rangle_t &\approx \sum_{j\neq 0} S[\Delta\Phi_{M,\Q} j] \left[\sum_{i}  \left<|\alpha_{0,i}(t)|^4\right>_t \right] \label{eq:variance structure factor sum}
\end{align}
where we made the far-field replacement $|\alpha_{0,i-{\rm LCM}(M,\Q)j}|^2\approx|\alpha_{0,i}|^2$.
Note that $\sum_{i} \left<|\alpha_{0,i}(t)|^4\right>_t$ is the inverse participation ratio (IPR) in the AA lattice. 

(3) for $\Delta\Phi \rightarrow 0$, we find 
\begin{align}
\langle{\rm var}\rangle_t 
=&  \sum_{j, i} \left<|\alpha_{0,i-{\rm LCM}(M,\Q)j}(t)|^2|\alpha_{0,i}(t)|^2\right>_t\nonumber\\
&\ \ \ \  - \sum_i \left<|\alpha_{0,i}(t)|^4\right>_t
\nonumber\\
\approx& \frac{1}{{\rm LCM}(M,\Q)} - \sum_{i} \left<|\alpha_{0,i}(t)|^4\right>_t \nonumber \\
& \rightarrow \frac{1}{{\rm LCM}(M,\Q)}
\label{eq: Thomae function}
\end{align}
where we assumed that the wavefunction is uniform on the scale of ${\rm LCM}(M,\Q)$ sites (a poor assumption in the insulating phase, and even in our data there is a $\approx2\times$ difference in typical averaged populations) and broke the first term into ${\rm LCM}(M,\Q)$ individual terms $|\alpha_{0,i-{\rm LCM}(M,\Q)j-m}(t)|^2|\alpha_{0,i}(t)|^2$ with $m$ from $0$ to ${\rm LCM}(M,\Q)-1$.
With that we introduced a new sum over $m$, that we used to complete the $j$ sum to be on every lattice site.
At which point we  used the normalization condition to do this double sum, and the $\rightarrow$ is assuming a spreading wavepacket for which the sum of the IPR $i$ will fall to zero.
%%%%%%%%%%%%%%%%%%%%%%%%%%%%%%%%%%%%%%%%%%%
%
\subsection{AA lattices in real space}
\label{Appendix: AA lattices in real space}
To best align with our our TOF data, we explained our observations in a momentum space picture. We note that an analogous description in real space can be derived using the dual AA model from that just presented, where the noise-suppression for irrational transverse flux is understood in terms of spatial self-averaging.
Inserting the Fourier expansion
\begin{equation}
\ket{m,n}=\frac{1}{\sqrt{M}}\sum_{q_m}e^{-i2\pi q_m m}\ket{q_m,n}
\label{eq: Fourier transform to spatial AA chain}
\end{equation} 
where $q_m\in \{0,\frac{1}{M},\cdots,\frac{M-1}{M}\}$, into the Hamiltonian in Eq.~\eqref{eq: HH Hamiltonian}, leads to 
\begin{equation} 
\begin{aligned}
&\HAA(\phibar-q_m)
\\
=&-J_{x}\sum_{n}(\ket{q_m,n+1}\bra{q_m,n}
+{\rm h.c.})
\\
&-2J_{s}\sum_{n}\cos[2\pi(\Phi n+\phibar-q_m)]\ket{q_m,n}\bra{q_m,n}
\label{eq:Spatial AA chain Hamiltonian}
\end{aligned}
\end{equation}
where $\hat{H}=\sum_{q_m}\HAA(\phibar-q_m)$. The original HH model is transformed into $M$ decoupled AA chains in real space, where the quasi-momentum $q_m$ shifts the phase of the corresponding chain.

Without repeating the analogous derivations as in momentum space, we simply point out that we reach the same results for the mean and variance of the time evolution. One key step of the derivation takes advantage of the displacement property of the Hamiltonian. 

The Hamiltonian Eq.~\eqref{eq:Spatial AA chain Hamiltonian} at longitudinal site $n$ is identical to the Hamiltonian at site $n=0$ with the Peierls phase factor changed to $\phibar'=\Phi n+\phibar$.
Expressed in terms of the displacement operator that shifts the longitudinal sites as $\ket{n+\delta n}=\hat{D}(\delta n)\ket{n}$, the displacement property of the Hamiltonian is
\begin{align}
\hat{D}(\delta n)\hat{H}(\phibar)\hat{D}^\dagger(\delta n)&=\hat{H}(\phibar-\Phi\delta n).
\label{eq: displacement operator}
\end{align}
At irrational fluxes $\Phi$, for a large system, summing over $n$ randomly samples the phases, equivalent to averaging over $\phibar$. Therefore, each individual time evolution is expected to evolve as the averaged evolution--a spatial self-averaging effect. On the other hand, at rational fluxes, averaging over $n$ only allows sampling $\Q$ different phases in each chain and a total of ${\rm LCM}(M,\Q)$ phases including all chains. For example, at $\P/\Q=2/3$ and $\phibar=0$, only three phases 0, $2\pi/3$ and $4\pi/3$ are sampled, inequivalent to averaging over $\phibar$. Thus, each time evolution is expected to have its own unique trajectory. 

\section{System size}

\begin{figure*}[t]
\begin{center}
\includegraphics[width=\textwidth]{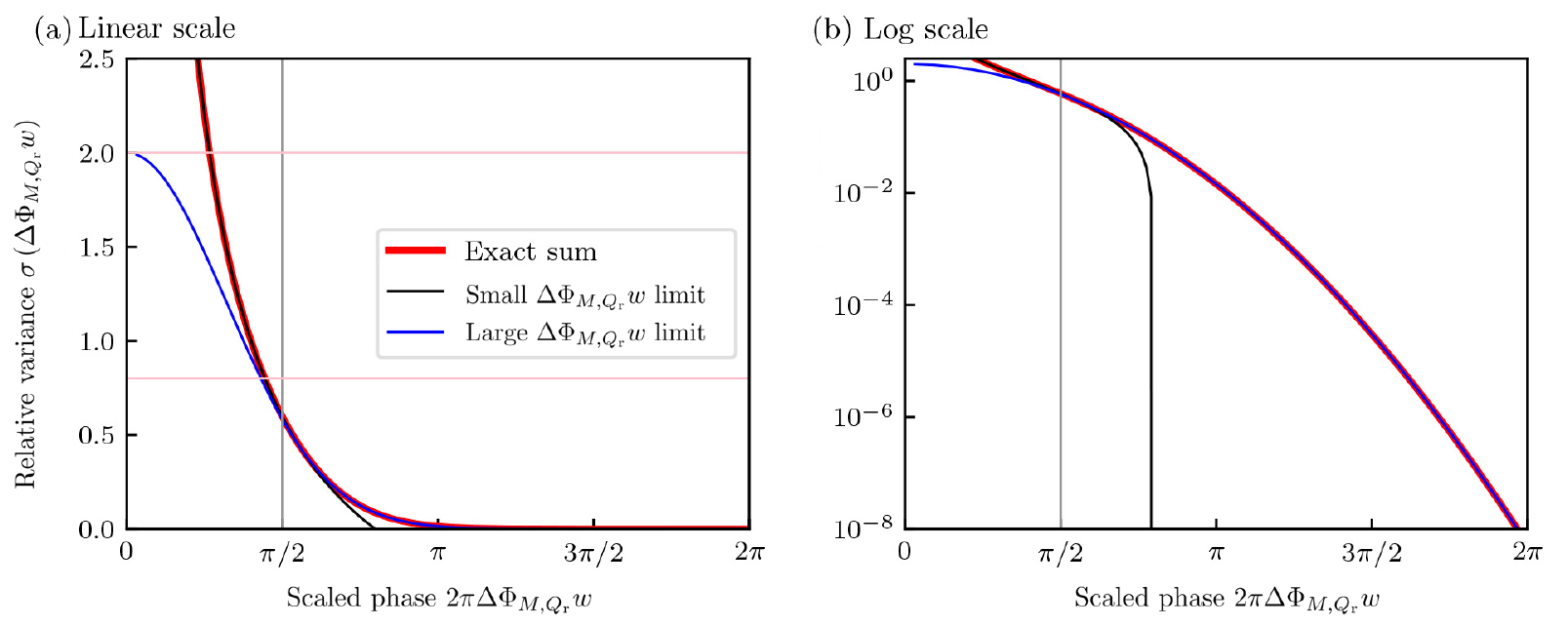}
\end{center}
\caption{\textbf{Time evolution uncertainty plotted in (a) linear and (b) log scale for a Gaussian wavefunction.} 
In both cases, the thick red curve comes from exactly evaluating Eq.~\eqref{eq:thetafunction}, while the black curve results from the small $w \Delta\Phi_{M,\Q}$ expansion and the black curve results from the large $w \Delta\Phi_{M,\Q}$ expansion.
The vertical grey line marks $2\pi w \Delta\Phi_{M,\Q} = \pi/2$ were we expect the validity of these expansions to crossover.
The pink lines mark the maximum possible noise for quantum states drawn completely at random: the top line at 2 is for $M\rightarrow\infty$, and the bottom line at 0.8 is for $M=3$.
}
\label{fig:gaussianpeak}
\end{figure*} 

We define the relative variance
\begin{align*}
\sigma(\Delta\Phi_{M,\Q}w) &= \frac{\langle{\rm var}\rangle_t}{\left\langle \sum_{\mf}\left\langle P_{\mf}(t)\right\rangle_\phi^2 \right\rangle_t}.
\end{align*}
Assuming an initial real space Gaussian wavepacket, 
\begin{align*}
\psi_0(n)&=\frac{1}{\pi^{1/4}w^{1/2}}\exp\left[-\frac{1}{2}\left(\frac{n}{w}\right)^2\right],
\end{align*}
with width $w$, from its momentum space counterpart, the static structure factor can be obtained by using Eq.~\eqref{eq:wavefunc overlap}
\begin{align*}
S(j)=\exp\left[-\frac{1}{2}\left(2\pi\Delta\Phi_{M,\Q} wj\right)^2\right]
\end{align*}
Along with Eq.~\eqref{eq:variance structure factor sum} and $\left\langle\sum_{\mf}\left\langle P_{\mf}(t)\right\rangle_\phi^2 \right\rangle_t\approx\sum_{i} \left<|\alpha_{0,i}(t)|^4\right>_t$, this allows us to directly compute the relative variance in the long-evolution time limit
\begin{align}
\sigma({\Delta\Phi_{M,\Q}}w)&=\vartheta_3\left[0,i2\pi\left(\Delta\Phi_{M,\Q} w\right)^2\right]-1 
\label{eq:thetafunction}
\end{align}
in terms of $\vartheta_3(z,\tau)=\sum_{j=-\infty}^{\infty}q^{j^2}\eta^j$ the third elliptic theta function, where $q=e^{i\pi\tau}$ and $\eta=e^{2\pi iz}$. 

For $|\Delta\Phi_{M,\Q}| w\gtrsim 1/4$, keeping only the $j=0,\pm1$ terms, the function is very well described by
\begin{align}
\sigma({\Delta\Phi_{M,\Q}}w) &\rightarrow 2 \exp\left[-\frac{1}{2} \left(2\pi\Delta\Phi_{M,\Q} w\right)^2\right] 
\label{eq:system size fit}
\end{align}

For $|\Delta\Phi_{M,Q}|w \lesssim 1/4$, using one of the Jacobi identities $\vartheta_3(z/\tau,-1/\tau)=\alpha\vartheta_3(z,\tau)$, where $\alpha=\sqrt{-i\tau}\exp\left[\pi iz^2/\tau\right]$ and keeping only the $j=0$ term, the function is very well described by
\begin{align*}
\sigma({\Delta\Phi_{M,\Q}}w) &\rightarrow \frac{\sqrt{2\pi}}{2\pi|\Delta\Phi_{M,\Q}| w} -1
\end{align*}
However, it is worth noting that this limit requires a very long evolution time to satisfy the approximation made in Eq.~\eqref{eq:variance structure factor sum}, from which we obtained $\sigma({\Delta\Phi_{M,\Q}}w)$.  

Figure~\ref{fig:gaussianpeak} demonstrates the agreement of these two approximate limits with the explicitly evaluated summation, and clearly marks the Gaussian wings and enhancement for small $\sigma({\Delta\Phi_{M,\Q}}w)$.

\bibliography{My_Library}

\end{document}